\documentclass[apjl]{emulateapj}
\usepackage{psfig,amsfonts,amsmath,graphicx,natbib,apjfonts}
\def\ra#1#2#3{#1$^{\rm h}$#2$^{\rm m}$#3$^{\rm s}$}
\def\dec#1#2#3{$#1^\circ#2'#3''$}
\def\swift{{\it Swift}}
\def\nod{\nodata}

\def\prince{1}
\def\ociw{2}

\begin{document}

\title{The Host Galaxies of Short-Duration Gamma-Ray Bursts:
Luminosities, Metallicities, and Star Formation Rates}

\author{
E.~Berger\altaffilmark{\prince,}\altaffilmark{\ociw}
}

\altaffiltext{\prince}{Princeton University Observatory, Princeton, NJ
08544}

\altaffiltext{\ociw}{Observatories of the Carnegie Institution of
Washington, 813 Santa Barbara Street, Pasadena, CA 91101}

\begin{abstract} 

The association of some short-duration gamma-ray bursts (GRBs) with
elliptical galaxies established that their progenitors, unlike those
of long GRBs, belong to an old stellar population.  However, the
majority of short GRBs appear to occur in star forming galaxies,
raising the possibility that some progenitors are related to recent
star formation activity.  Here we present optical spectroscopy of
these hosts and measure their luminosities, star formation rates, and
metallicities.  We find luminosities of $L_B\approx 0.1-1.5$ $L_*$,
star formation rates of ${\rm SFR}\approx 0.2-6$ M$_\odot$ yr$^{-1}$,
and metallicities of $12+{\rm log(O/H)}\approx 8.5-8.9$ ($Z\approx 
0.6-1.6$ $Z_\odot$).  A detailed comparison to the hosts of long GRBs 
reveals systematically higher
luminosities, lower specific star formation rates (${\rm SFR}/L_B$) by
about an order of magnitude, and higher metallicities by about 0.6
dex.  The K-S probability that the short and long GRB hosts are drawn
from the same underlying galaxy distribution is only $\sim 10^{-3}$.  
On the other hand, short GRB hosts exhibit excellent agreement with the
specific star formation rates and the luminosity-metallicity relation 
of field galaxies at $z\sim 0.1-1$.  We thus
conclude that short GRB hosts are not dominated by young stellar
populations like long GRBs hosts.  Instead, short GRB hosts appear to
be drawn uniformly from the underlying galaxy distribution, indicating
that the progenitors have a wide age distribution of several Gyr.

\end{abstract}


\section{Introduction}
\label{sec:intro}

The properties of gamma-ray burst (GRB) host galaxies provide
important insight into the nature of the burst progenitors.  In the
case of the long-duration GRBs ($T_{90}\gtrsim 2$ s), the hosts are
blue star forming galaxies with high specific star formation rates and
sub-$L_*$ luminosities (e.g.,
\citealt{hf99,cba02,bck+03,ldm+03,chg04,sgl08}).  Moreover, the bursts
themselves trace the star formation activity within their hosts
\citep{bkd02,fls+06}.  At low redshift, $z\lesssim 0.3$, long GRBs
appear to occur preferentially in low luminosity and metallicity
galaxies \citep{sgb+06,sgl08}, although it remains unclear whether this 
is a causal connection or a byproduct of the intense star formation
activity \citep{bfk+07}.  Taken together, these properties provided an
early indication that the progenitors of long GRBs are massive stars,
and ultimately they may shed light on the conditions (if any) that
favor the formation of the progenitors.

The more recent discovery of afterglow emission from short-duration
GRBs led to associations with both star forming and elliptical host
galaxies \citep{bpc+05,ffp+05,gso+05,bpp+06}.  The association with
elliptical galaxies demonstrates unambiguously that the progenitors of
at least some short GRBs are related to an old stellar population,
consistent with the popular model of compact object mergers (NS-NS or
NS-BH; e.g., \citealt{elp+89,npp92}).  In addition, {\it Hubble Space
Telescope} imaging of the star forming host galaxy of GRB\,050709
revealed that, unlike in the case of the long GRBs, the burst was not
associated with a region of active star formation \citep{ffp+05}.

Subsequent to the discovery of the first short GRB afterglows and 
hosts, we have shown that a substantial fraction of these events 
($1/3-2/3$) reside at higher redshifts than previously suspected, 
$z\gtrsim 0.7$ \citep{bfp+07,cbn+08}.  Spectroscopic observations 
indicate that a
substantial fraction of these galaxies are undergoing active star
formation.  Indeed, of the current sample of short GRBs localized to
better than a few arcseconds (23 bursts), $\approx 45\%$ reside in
star forming galaxies compared to only $\approx 10\%$ in elliptical
galaxies\footnotemark\footnotetext{The other $\approx 45\%$ remain
currently unclassified due to their faintness, a lack of obvious
spectroscopic features, or the absence of deep follow-up.}.  This
result raises the question of whether some short GRBs are related to
star formation activity rather than an old stellar population, and if
so, whether the star formation properties are similar to those in long
GRB host galaxies.  The answer will shed light on the diversity of
short GRB progenitors, in particular their age distribution and the
possibility of multiple progenitor populations.

Here we present optical spectroscopy of short GRB host galaxies and
measure their luminosities, metallicities, and star formation rates
(\S\ref{sec:obs} and \S\ref{sec:prop}).  We then assess their specific
star formation rates and luminosity-metallicity relation, and compare
these results with the properties of long GRB hosts and field star
forming galaxies (\S\ref{sec:disc}).  Finally, we use these
comparisons to draw conclusions about the progenitor population, and
we outline future host galaxy studies that will provide continued
constraints on the progenitors (\S\ref{sec:conc}).  Throughout the
paper we use the standard cosmological parameters, $H_0=70$ km
s$^{-1}$ Mpc$^{-1}$, $\Omega_m=0.27$, and $\Omega_\Lambda=0.73$.

\section{Observations}
\label{sec:obs}

\begin{figure}
\epsscale{1}
\plotone{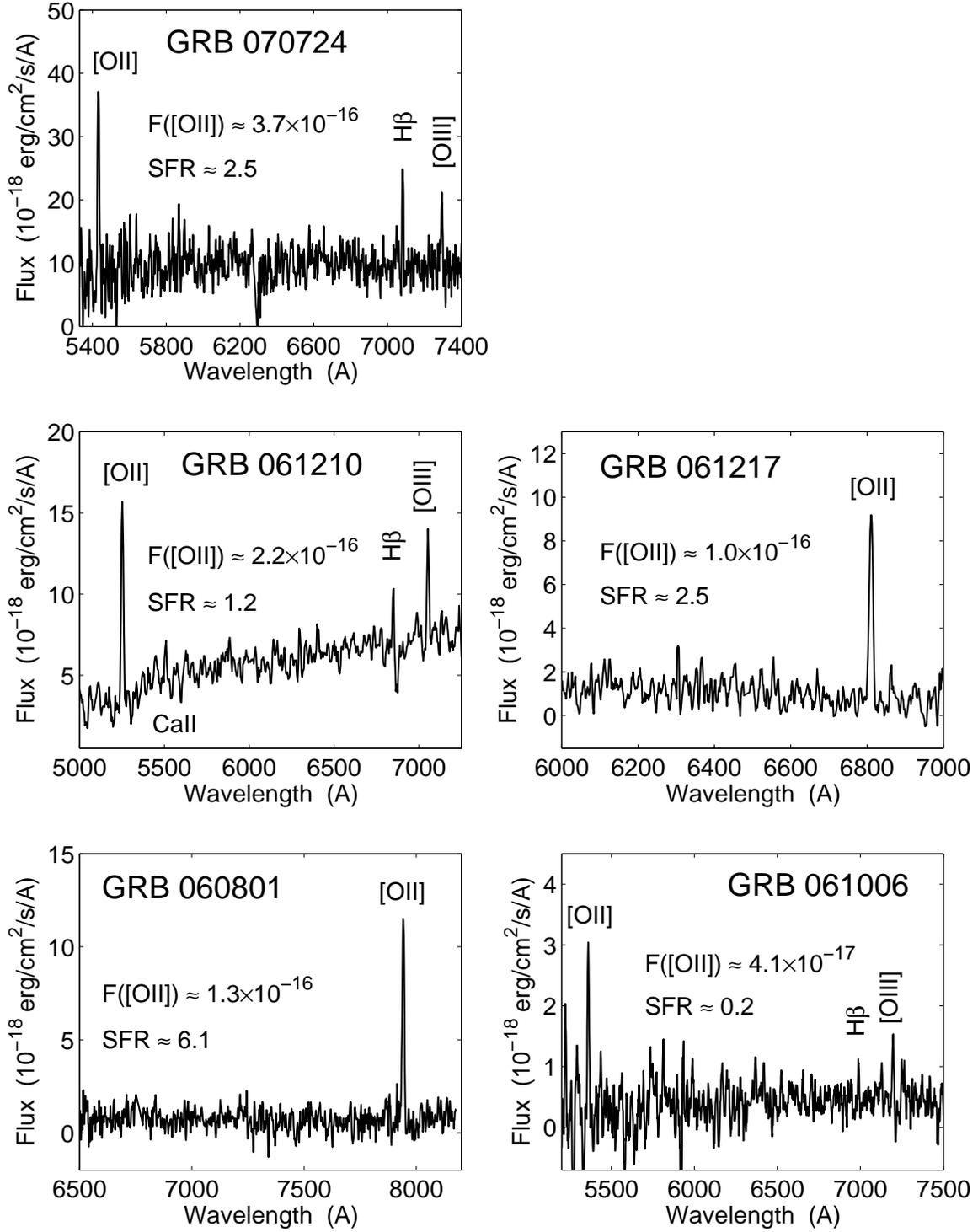}
\caption{Optical spectra of short GRB host galaxies obtained with
Gemini/GMOS and Magellan/LDSS3.  For details see \S\ref{sec:obs} and
\citet{bfp+07}.  The relevant emission lines are marked, and star
formation rates from the ${\rm [OII]}\lambda 3727$ doublet are
provided.
\label{fig:spectra}}
\end{figure}

In Table~\ref{tab:grbs} we summarize the properties of all short GRBs
with \swift\ X-ray Telescope (XRT) localizations, and their host
galaxies.  All references for the burst and host properties are
provided in Table~\ref{tab:grbs}, including in particular
\citet{bfp+07} and \citet{cbn+08} for the details of our spectroscopic
observations.  We present the flux-calibrated spectra of the host
galaxies of GRBs 060801, 061006, 061210, 061217, and 070724 in
Figure~\ref{fig:spectra}.  Spectra of the hosts of GRBs 070429b and
070714b are presented in \citet{cbn+08}, while spectra of the hosts of
GRBs 050709 and 051221a are provided in \citet{ffp+05} and
\citet{sbk+06}, respectively.  Below we summarize our new
spectroscopic observations of GRB\,070724.

\subsection{GRB\,070724}

A putative host galaxy was identified within the $2.2''$ radius XRT
error circle of this short burst in archival Digital Sky Survey images
\citep{gcn6661}.  No coincident variable optical, near-IR, or radio
source was detected \citep{gcn6662,gcn6665,gcn6666,gcn6667}.

We obtained imaging and spectroscopy of the likely host galaxy using
the Gemini Multi-Object Spectrograph (GMOS; \citealt{hja+04}) mounted
on the Gemini-South 8-m telescope.  We confirm the presence of the
galaxy and measure its brightness to be $r_{\rm AB}=20.56\pm 0.03$
mag.  The probability of chance coincidence for a galaxy of this
brightness within the XRT error circle is only $8\times 10^{-3}$
\citep{bsk+06}.

The spectroscopic observations lasted a total of 3600 s, using the
GMOS R400 grating at central wavelengths of 7000 and 7050 \AA\ with a
$1''$ slit (resolution of about 7 \AA).  The data were reduced using
the {\tt gemini} package in IRAF.  Wavelength calibration was
performed using CuAr arc lamps and air-to-vacuum and heliocentric
corrections were applied.  We identify several emission lines
corresponding to ${\rm [OII]}\lambda 3727$, ${\rm [OIII]}\lambda
\lambda 4959,5007$, and H$\beta$ at a redshift of $z=0.4571\pm 0.0003$
(Figure~\ref{fig:spectra}).

\section{Host Galaxy Properties}
\label{sec:prop}

The sample of short GRBs presented in this paper is comprised of all
events with \swift/XRT positions (23 bursts), with typical
uncertainties of $\sim 2-5''$ \citep{but07,gtb+07}.  Of these events,
nine bursts have been further localized to sub-arcsecond precision
based on detections in the optical, near-IR, radio, and/or X-rays.  In
eight of these nine cases a host galaxy has been
identified\footnotemark\footnotetext{The single exception is
GRB\,070707 for which our deepest limit is $r\gtrsim 24.4$ mag.}, and
in six cases its redshift has been measured.  These are the most
secure host galaxy associations in the sample, with a redshift range
of $z\approx 0.16-0.92$, an optical magnitude range of $r\approx
18-25$ mag, and a ratio of late- to early-type of $5\!:\!1$ (hereafter,
{\it Sample 1}).  The additional two bursts with sub-arcsecond
positions and no current redshift measurements have faint hosts with 
$r=24.8$ and 26.3 mag.  As discussed in \citet{bfp+07}, these events 
are likely located at $z\gtrsim 0.7$.

Of the 14 short bursts with only XRT positions, 12 have deep follow-up
observations that led to the identification of galaxy counterparts,
with $r\approx 17-26$ mag; the chance coincidence probabilities for
these galaxy associations range from $\sim 10^{-3}$ to 0.1 (for a 
detailed discussion see \citealt{bfp+07}).  Six of these 12 hosts have 
spectroscopically
measured redshifts, $z\approx 0.23-1.13$ (hereafter, {\it Sample 2}).
These are the brighter hosts, $r\approx 17-23$ mag, and their 
probabilities of chance coincidence are thus the lowest \citep{bfp+07}.
Their ratio of late- to early-type is again $5\!:\!1$.  The remaining, 
fainter hosts likely reside at $z\gtrsim 0.7$ \citep{bfp+07}.

We focus here on the two samples of host galaxies with measured
spectroscopic redshifts.  As we stress below, our results remain
unaffected even if we use only the events with sub-arcsecond positions
({\it Sample 1}).

\subsection{Luminosities}

We infer the host galaxy absolute magnitudes in the rest-frame
$B$-band, $M_B$, using their observed $r$-band magnitudes
(Table~\ref{tab:grbs}).  At $z\lesssim 0.5$ the observed $r$-band
directly samples the rest-frame $B$-band, but for the hosts at $z\sim
1$ it provides a measure of the rest-frame $U$-band.  To transform the
latter to the $B$-band we use the typical rest-frame $U-B$ colors of
blue galaxies, appropriate for our star forming galaxy sample, as
measured from the DEEP2 galaxy survey, $U-B\approx 0.75$ mag
\citep{cnc+08}.  The observed dispersion in $U-B$ color results in a
$\sim 30\%$ uncertainty in $M_B$.

\begin{figure}
\epsscale{1}
\plotone{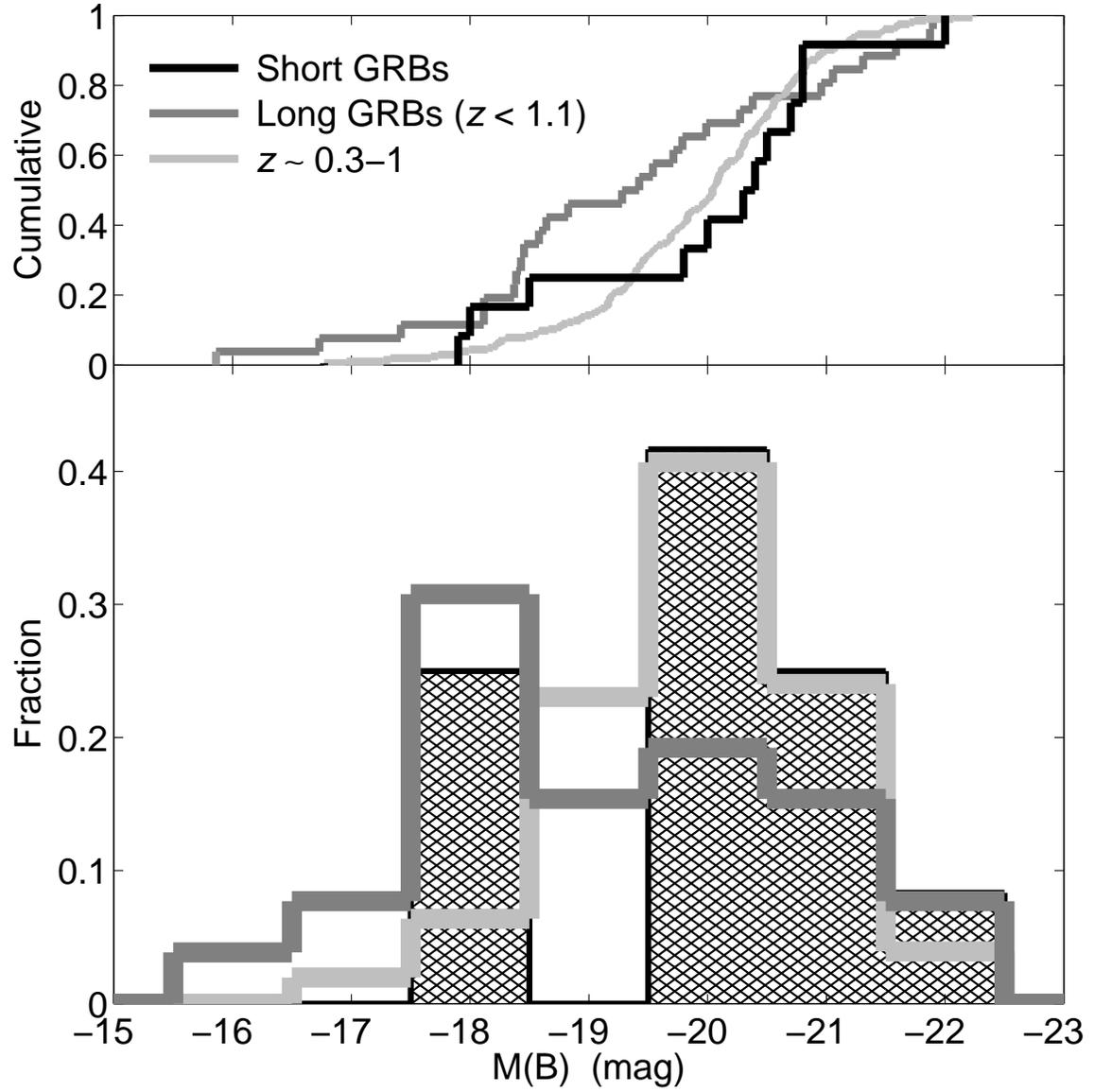}
\caption{Distribution of $B$-band absolute magnitudes for the hosts of
short (black) and long (dark gray) GRBs, as well as field star forming
galaxies from the GOODS-N survey (light gray; \citealt{kk04}).  A K-S
test of the cumulative distributions (top) indicates a probability of
only $10\%$ that the short GRB hosts are drawn from the same
distribution of long GRB hosts, and $60\%$ that they are drawn from
the field galaxy distribution.
\label{fig:mb}}
\end{figure}

For the hosts in {\it Sample 1} we find $M_B\approx -18$ to $-21$ mag,
or $L_B\approx 0.1-1.5$ $L_*$; the highest luminosity belongs to the
elliptical host galaxy of GRB\,050724 \citep{bpc+05}.  We use the
appropriate value of $L_*$ as a function of redshift determined from
the DEEP2 survey \citep{wfk+06}.  For the star forming hosts in {\it
Sample 2} we find $M_B\approx -20$ to $-21$ mag, or $L_B\approx 0.4-
1.4$ $L_*$; the elliptical host galaxy of GRB\,050509b has $L_B\approx
5$ $L_*$ \citep{bpp+06}.  The distribution of $M_B$ values is shown in
Figure~\ref{fig:mb}.

\subsection{Star Formation rates}

We next use the ${\rm [OII]}\lambda 3727$ line luminosities to infer
the host star formation rates (Figure~\ref{fig:spectra}).  We use the
standard conversion, ${\rm SFR}=(1.4\pm 0.4)\times 10^{-41}\, L_{\rm
[OII]}$ M$_\odot$ yr$^{-1}$ \citep{ken98}.  For the star forming hosts
in {\it Sample 1} we find ${\rm SFR}\approx 0.2-1$ M$_\odot$
yr$^{-1}$, while the elliptical host of GRB\,050724 has ${\rm SFR}
\lesssim 0.05$ M$_\odot$ yr$^{-1}$ \citep{bpc+05}.  For the galaxies
in {\it Sample 2} we infer ${\rm SFR}\approx 1-6$ M$_\odot$ yr$^{-1}$,
with a limit of $\lesssim 0.1$ M$_\odot$ yr$^{-1}$ for the elliptical
host of GRB\,050509b \citep{bpp+06}.  The star formation rate for each
host galaxy is provided in Table~\ref{tab:grbs}.

Using the absolute magnitudes inferred above, we find that the
specific star formation rates are ${\rm SFR}/L_B\approx 1-10$
M$_\odot$ yr$^{-1}$ $L_*^{-1}$ for the star forming hosts; see
Figure~\ref{fig:ssfr}.  For the two elliptical host galaxies the upper
limits are ${\rm SFR}/L_B\lesssim 0.03$ M$_\odot$ yr$^{-1}$
$L_*^{-1}$.

\subsection{Metallicities}
\label{sec:metal}

For five of the twelve host galaxies we have sufficient spectral
information to measure the metallicity\footnotemark\footnotetext{The
relevant emission lines are ${\rm [OII]}\lambda 3727$, H$\beta$, ${\rm
[OIII]}\lambda \lambda 4959,5007$, H$\alpha$, and ${\rm [NII]}\lambda
6584$.}.  We use the standard metallicity diagnostics, $R_{23}\equiv
(F_{\rm [OII]\lambda 3727}+F_{\rm [OIII]\lambda\lambda 4959,5007})
/F_{\rm H\beta}$ \citep{peb+79,kk04} and $F_{\rm [NII]\lambda
6584}/F_{\rm H\alpha}$.  The value of $R_{23}$ depends on both the
metallicity and ionization state of the gas, which we determine using
the ratio of oxygen lines, $O_{32}\equiv F_{\rm [OIII]\lambda\lambda
4959,5007}/F_{\rm [OII] \lambda 3727}$.

We note that the $R_{23}$ diagnostic is double-valued with low and
high metallicity branches (e.g., \citealt{kd02}).  This degeneracy can
be broken using ${\rm [NII]}/{\rm H}\alpha$ when these lines are
accessible.  To facilitate a subsequent comparison with field galaxies
(\S\ref{sec:disc}) we use the $R_{23}$, $O_{32}$, and ${\rm
[NII]}/{\rm H}\alpha$ calibrations of \citet{kk04} (their Equations 12
and 18).  We note that the typical uncertainty inherent in the
calibrations is about 0.1 dex.   

The line fluxes for the five short GRB hosts, and the resulting values
of $R_{23}$ and $O_{32}$ are provided in Table~\ref{tab:lines}.  We 
adopt the solar metallicity from \citet{ags05}, $12+{\rm log(O/H)}=
8.66$.  For
the host of GRB\,061006 we find $12+{\rm log(O/H)}\approx 8.6$ for the
upper $R_{23}$ branch and $\approx 8.0-8.5$ for the lower branch.  For
the host of GRB\,070724 we find $12+{\rm log(O/H)}\approx 8.9$ for the
upper branch, and $\approx 7.6-8.1$ for the lower branch.  We find a
similar range of values for the host of GRB\,061210, but the ratio 
$F_{\rm [NII]}/F_{\rm H\alpha}\approx 0.2$, indicates $12+{\rm 
log(O/H)} \gtrsim 8.6$, thereby breaking the degeneracy and leading to 
the upper branch solution, $12+{\rm log(O/H)}\approx 8.9$.

For the previously-published host of GRB\,051221a we use the line
fluxes provided in \citet{sbk+06}, and derive similar values to those
for the host of GRB\,070724.  Finally, for the host galaxy of
GRB\,050709 we lack a measurement of the ${\rm [OII]}$ emission line,
and we thus rely on ${\rm [NII]}/{\rm H}\alpha$ to
infer\footnotemark\footnotetext{We note that \citet{pbc+06} infer a
metallicity of $12+{\rm log(O/H)}\approx 8.2$ for GRB\,050709, but
their value is based on a different calibration of ${\rm [NII]}/{\rm
H}\alpha$.  Our inferred value allows for a self-consistent comparison
with field galaxies and long GRB hosts.} $12+{\rm log(O/H)}\approx
8.5$.  The dominant source of uncertainty in this measurement is the
unknown value of $O_{32}$, but using a spread of a full order of
magnitude results in a metallicity uncertainty of 0.2 dex.

For the hosts with double-valued metallicities (GRBs 051221a, 061006,
and 070724) we follow the conclusion for field galaxies of similar
luminosities and redshifts that the appropriate values are those for
the $R_{23}$ upper branch \citep{kk04}.  This conclusions was
advocated by \citet{kk04} based on galaxies in their sample with
measurements of both $R_{23}$ and ${\rm [NII]/H}\alpha$.  It is
similarly supported by our inference for the host galaxy of
GRB\,061210.  The metallicities as a function of host luminosity are
shown in Figure~\ref{fig:lz}.

\section{A Comparison to Long GRB Host Galaxies and Field Galaxies}
\label{sec:disc}

To place the host galaxies of short GRBs in a broader context we now
turn to a comparison of their properties with those of long GRB hosts
and field star forming galaxies at a similar redshift range.  For the
field sample we use the $\approx 200$ emission line galaxies at
$z\approx 0.3-1$ from the Great Observatories Origins Deep
Survey-North (GOODS-N) field studied spectroscopically as part of the
Team Keck Redshift Survey \citep{kk04}.  We select this sample due to
its large size, availability of luminosity, metallicity, and star
formation rate measurements, and because its limiting magnitude for
spectroscopy is similar to that for the short GRB host galaxies.

A comparison of the $B$-band absolute magnitude distributions of short
and long GRB hosts in the same redshift range ($z\lesssim 1.1$) is
shown in Figure~\ref{fig:mb}.  The long GRB hosts range from $M_B
\approx -15.9$ to $-21.9$ mag, with a median value of $\langle M_B
\rangle\approx -19.2$ mag ($\langle L_B\rangle\approx 0.2$ $L_*$;
\citep{bfk+07}).  Thus, the long GRB hosts extend to lower
luminosities than the short GRB hosts, with a median value that is
about 1.1 mag fainter.  A Kolmogorov-Smirnov (K-S) test indicates that
the probability that the short and long GRB hosts are drawn from the
same underlying distribution is $0.1$.  On the other hand, a
comparison to the GOODS-N sample reveals a similar distribution, and
the K-S probability that the short GRB hosts are drawn from the field
sample is 0.6.

\begin{figure}
\epsscale{1}
\plotone{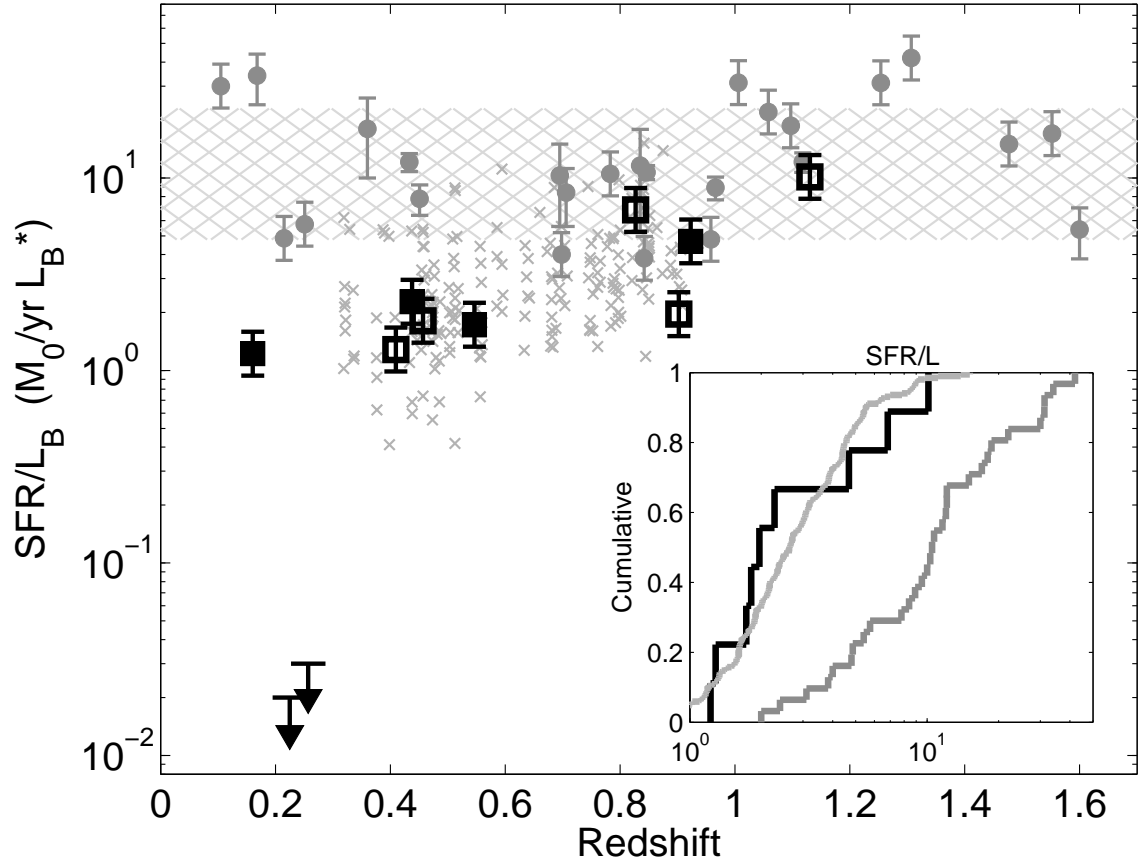}
\caption{Specific star formation rates as a function of redshift for
the host galaxies of short (black) and long (gray) GRBs, as well as
field galaxies from the GOODS-N survey (crosses; \citealt{kk04}).
Solid and open symbols designate short GRBs with sub-arcsecond
positions and XRT positions only, respectively.  Upper limits for the
elliptical hosts of GRBs 050509b and 050724 are also shown.  The
cross-hatched region marks the median and standard deviation for the
long GRB host sample.  The inset shows the cumulative distributions
for the three samples.  The K-S probability that the short and long
GRB hosts are drawn from the same distribution is only $0.3\%$, while
the strong overlap with the field sample leads to a K-S probability of
$60\%$.
\label{fig:ssfr}} 
\end{figure}

\begin{figure}
\epsscale{1}
\plotone{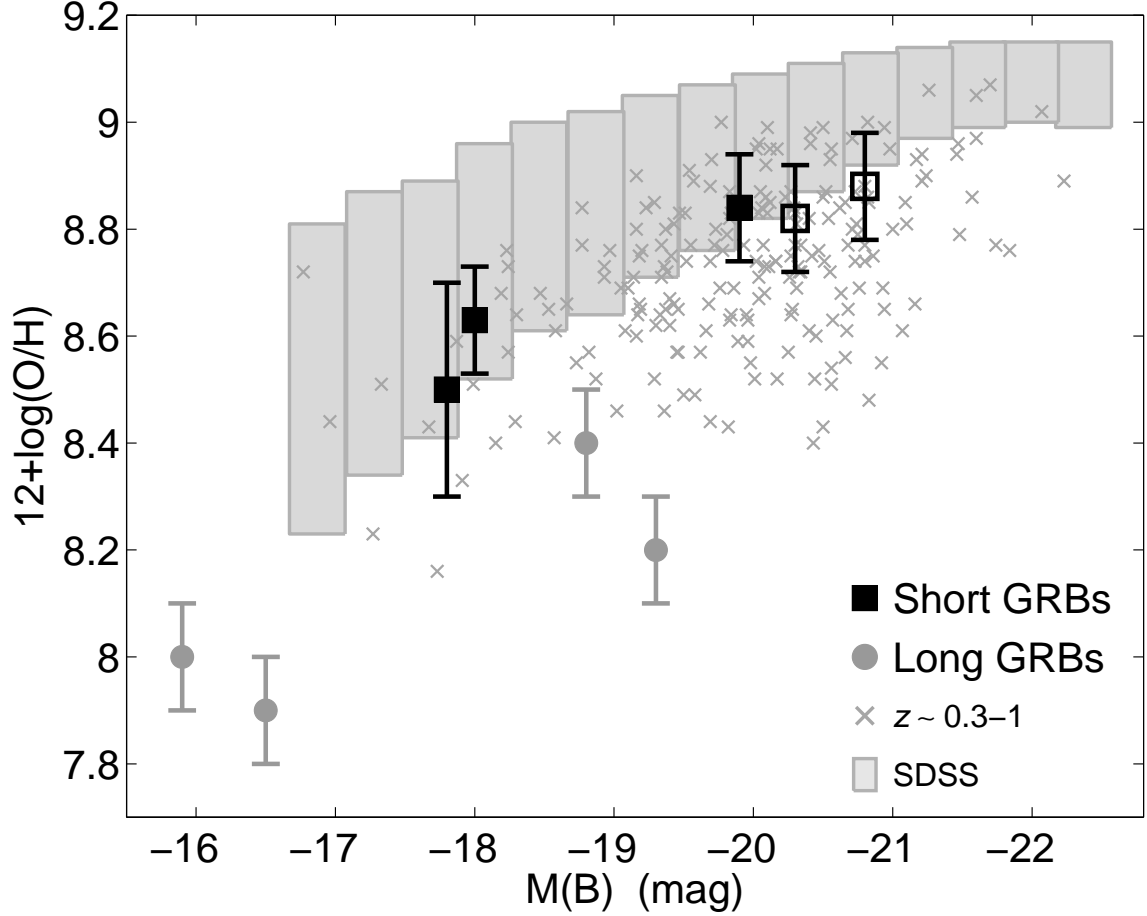}
\caption{Metallicity as a function of $B$-band absolute magnitude for
the host galaxies of short (black) and long (gray) GRBs.  The gray
bars mark the $14-86$ percentile range for galaxies at $z\sim 0.1$
from the Sloan Digital Sky Survey \citep{thk+04}, while crosses
designate the same field galaxies at $z\sim 0.3-1$ shown in
Figure~\ref{fig:ssfr} \citep{kk04}.  Both field samples exhibit a
clear luminosity-metallicity relation.  The long GRB hosts tend to
exhibit lower than expected metallicities \citep{sgb+06}, while the
hosts of short GRBs have higher metallicities by about 0.6 dex, are
moreover in excellent agreement with the luminosity-metallicity
relation.
\label{fig:lz}} 
\end{figure}

We reach a similar conclusion based on a comparison of specific star
formation rates.  For long GRB hosts the inferred star formation rates
range from about 0.2 to 50 M$_\odot$ yr$^{-1}$, and their specific
star formation rates are about $3-40$ M$_\odot$ yr$^{-1}$ $L_*^{-1}$,
with a median value of about 10 M$_\odot$ yr$^{-1}$ $L_*^{-1}$
\citep{chg04}.  As shown in Figure~\ref{fig:ssfr}, the specific star
formation rates of short GRB hosts are systematically lower than those
of long GRB hosts, with a median value that is nearly an order of
magnitude lower.  Indeed, the K-S probability that the short and long
GRB hosts are drawn from the same underlying distribution is only
$3.5\times 10^{-3}$.  This is clearly seen from the cumulative 
distributions of specific star formation rates for each sample 
(inset of Figure~\ref{fig:ssfr}).

On the other hand, a comparison to the specific star formation rates
of the GOODS-N field galaxies reveals excellent agreement
(Figure~\ref{fig:ssfr}).  The K-S probability that the short GRB hosts
are drawn from the field galaxy distribution is 0.6.  This result
remains unchanged even if we use only the host galaxies in {\it Sample 1}.
Thus, short GRB hosts are drawn from the normal population of star
forming galaxies at $z\lesssim 1$, in contrast to long GRB hosts,
which have elevated specific star formation rates, likely as a result
of preferentially young starburst populations \citep{chg04,sgl08}.

Finally, the metallicities measured for short GRB hosts are in
excellent agreement with the luminosity-metallicity relation for field
galaxies at $z\sim 0.1-1$ (Figure~\ref{fig:lz};
\citealt{kk04,thk+04}).  The two hosts with $M_B\approx -18$ mag have
$12+{\rm log(O/H)}\approx 8.6$, while those with $M_B\approx -20$ to
$-21$ mag have $12+{\rm log(O/H)}\approx 8.8-8.9$, following the
general trend.  On the other hand, the short GRB host metallicities
are systematically higher than those of long GRB hosts, which have
been argued to have lower than expected metallicities \citep{sgb+06}.
The median metallicity of short GRB hosts is about 0.6 dex higher than
for long GRB hosts, and there is essentially no overlap between the
two host populations.

\section{Summary and Conclusions}
\label{sec:conc}

We present optical spectroscopy of several short GRB host galaxies,
and a complete compilation of all short GRBs with \swift/XRT positions
(23 bursts).  About one half of the sample has spectroscopically
identified host galaxies, with $z\approx 0.1-1.1$.  The ratio of star
forming to elliptical galaxies in this spectroscopic sample is 
$5\!:\!1$, regardless
of whether we use only the objects with sub-arcsecond positions, or
include those with XRT positions.  We note that the maximum allowed 
fraction
of elliptical galaxies in the \swift/XRT sample, assuming that all of
the currently-unclassified hosts turn out to be ellipticals, is
$55\%$.  In the formulation of \citet{zr07}, with a power law age
distribution, $P(\tau)\propto\tau^n$ (where $\tau$ is the time delay
between formation and merger),
the range of early- to late-type ratios allowed by the data leads to
$n\lesssim 1$; if the ratio is indeed $\sim 20\%$ then $n\lesssim -1$.

Despite the fact that most short GRBs occur in star forming galaxies,
their properties are strongly distinct from those of long GRB hosts.  The
rest-frame $B$-band luminosity distribution of the short GRB hosts is
systematically brighter than for long GRB hosts in the same redshift
range.  An even stronger difference is apparent in the specific star
formation rates, with a median value for short GRB hosts that is
nearly an order of magnitude lower than for long GRB hosts.
Similarly, the metallicities of the short GRB hosts are about 0.6 dex
higher than those of long GRB hosts, and unlike the long GRB hosts
they follow the luminosity-metallicity relation of field galaxies.  To
the extent that the mean properties of the host galaxies reflect the
identity of the progenitors, this clearly indicates that the
progenitors of long and short GRBs are themselves distinct, supporting
additional lines of evidence such as the lack of supernova
associations in short GRBs.

On the other hand, a comparison to a large sample of star forming
field galaxies in a similar redshift range reveals excellent agreement
in terms of specific star formation rates and the
luminosity-metallicity relation.  Indeed, the K-S probability that the
short GRB hosts are drawn from the field galaxy population is high,
$\approx 0.6$.  Thus, short GRBs select galaxies that are
representative of the average stellar populations at least to $z\sim
1$.

These comparisons, along with the presence of some short GRBs in
elliptical galaxies, indicate that the progenitor ages span a wide
range, $\sim 0.1-10$ Gyr.  However, the overall dissimilarity to the
hosts of long GRBs, which appear to be dominated by young stellar
populations ($\lesssim 0.1$ Gyr; \citealt{chg04}), indicates that only
a small fraction of short GRBs ($\lesssim 1/3$) are likely to arise
from a prompt population of progenitors.  These conclusions provide
additional support and constraints on the binary coalescence model of
short GRBs, but are at odds with young progenitor populations such as
magnetars.

We finally note that while nearly half of the short GRB hosts remain
unclassified at present, the overall qualitative conclusion of our
study -- that short and long GRB hosts and hence the progenitors have
a different distribution of properties -- is robust.  In particular,
even if future spectroscopic observations of the unclassified hosts
reveal that they are more similar to those of long GRBs, the overall
dispersion in short GRB host properties will still be significantly
larger than for long GRB hosts, both in terms of metallicities and
specific star formation rates.  Naturally, any population of short GRB
hosts that is found to be similar to long GRB hosts may lead to the
conclusion that some of the progenitors are related to a young stellar
population (e.g., promptly merging binaries or magnetars); to
reiterate, our current limit on such a population is $\lesssim 1/3$.

Looking forward, we expect to advance our understanding of the
progenitors from three primary lines of host galaxy investigations.
First, with a growing sample of short GRB hosts we will be able to
assess whether the progenitor population is uniform, or perhaps
exhibits a bimodal distribution, with a contribution from prompt
(proportional to star formation) and delayed (proportional to stellar
mass) components, as appears to be the case for type Ia supernovae
(e.g., \citealt{slp+06}).  We expect the sample to grow both from new
events, and from continued optical and near-IR spectroscopy of the
existing hosts.  Second, high angular resolution imaging with the {\it
Hubble Space Telescope} can be used to investigate in detail the
location of short GRBs within their host galaxies.  This has already
been done for GRB\,050709, indicating that, unlike for long GRBs, the
burst was not associated with a region of active star formation
\citep{ffp+05}.  Finally, absorption spectroscopy of short GRB
afterglows, still unavailable in the present sample, will directly
reveal the type of environment (disk, halo, intergalactic medium) in
which the bursts explode.  Taken together, these studies promise to
shed light on the short GRB progenitor population(s), at least until
the advent of sensitive gravitational wave detectors in the next
decade.

\acknowledgements

Based in part on observations obtained at the Gemini Observatory,
which is operated by the Association of Universities for Research in
Astronomy, Inc., under a cooperative agreement with the NSF on behalf
of the Gemini partnership: the National Science Foundation (United
States), the Particle Physics and Astronomy Research Council (United
Kingdom), the National Research Council (Canada), CONICYT (Chile), the
Australian Research Council (Australia), CNPq (Brazil) and CONICET
(Argentina)

%\bibliographystyle{apj}
%\bibliography{journals_apj,refs,refs2,refs3}

\clearpage
\begin{deluxetable}{lcccccccccccl}
\tabletypesize{\footnotesize}
\tablecolumns{13}
\tabcolsep0in\footnotesize
\tablewidth{0pc}
\tablecaption{Short GRB and Host Galaxy Properties
\label{tab:grbs}}
\tablehead {
\colhead {GRB}             &
\colhead {$T_{90}$}        &
\colhead {$F_\gamma$}      &
\colhead {RA$\,^a$}        &
\colhead {Dec.}            &
\colhead {Uncert.}         &
\colhead {OA?}             &
\colhead {$z$}             &
\colhead {$R\,^b$}         &
\colhead {$L_B$}           &
\colhead {SFR}             &
\colhead {$12+{\rm log}({\rm O/H})$} &
\colhead {Refs.}           \\
\colhead {}                &
\colhead {(s)}             &
\colhead {(erg cm$^{-2}$)} &
\colhead {(J2000)}         &
\colhead {(J2000)}         &
\colhead {($''$)}          &
\colhead {}                &
\colhead {}                &
\colhead {(mag)}           &
\colhead {($L_*$)}         &
\colhead {(M$_\odot$/yr)}  &
\colhead {}                &
\colhead {}          
}
\startdata
050509b & 0.04    & $(9.5\pm 2.5)\times 10^{-9}$ & \ra{12}{36}{13.58} & \dec{+28}{59}{01.3} & 9.3 & N & 0.226  & $16.75\pm 0.05$ & 5    & $<0.1$  & \nod & 1--2      \\
050709  & 0.07    & $(2.9\pm 0.4)\times 10^{-7}$ & \ra{23}{01}{26.96} & \dec{-38}{58}{39.5} & 0.4 & Y & 0.1606 & $21.05\pm 0.07$ & 0.1  & 0.2     & 8.5  & 3--5      \\
050724  & 3.0$^c$ & $(3.9\pm 1.0)\times 10^{-7}$ & \ra{16}{24}{44.36} & \dec{-27}{32}{27.5} & 0.5 & Y & 0.257  & $18.19\pm 0.03$ & 1    & $<0.05$ & \nod & 6--8      \\
050813  & 0.6     & $(1.2\pm 0.5)\times 10^{-7}$ & \ra{16}{07}{57.19} & \dec{+11}{14}{57.8} & 3.8 & N & \nod$^d$ & \nod          & \nod & \nod    & \nod & 9--11     \\
051210  & 1.27    & $(8.1\pm 1.4)\times 10^{-8}$ & \ra{22}{00}{41.26} & \dec{-57}{36}{46.5} & 2.9 & N & \nod   & $23.80\pm 0.15$ & \nod & \nod    & \nod & 12--13    \\
051221a & 1.40    & $(1.2\pm 0.1)\times 10^{-6}$ & \ra{21}{54}{48.62} & \dec{+16}{53}{27.2} & 0.2 & Y & 0.5465 & $21.81\pm 0.09$ & 0.3  & 1.0     & 8.8  & 14--15    \\
%051227 & 8.0     & $(2.3\pm 0.3)\times 10^{-7}$ & \ra{08}{20}{58.06} & \dec{+31}{55}{34.2} & 2.7 & Y & \nod   & $25.49\pm 0.15$ & \nod & \nod    & \nod & 13,16--17 \\
%	     	       			    								                   		       
060121  & 1.97    & $(4.7\pm 0.4)\times 10^{-6}$ & \ra{09}{09}{51.99} & \dec{+45}{39}{45.6} & 0.1 & Y & \nod   & $26.26\pm 0.30$ & \nod & \nod    & \nod & 13,16--17 \\
060313  & 0.70    & $(1.1\pm 0.1)\times 10^{-6}$ & \ra{04}{26}{28.42} & \dec{-10}{50}{39.9} & 0.2 & Y & \nod   & $24.83\pm 0.20$ & \nod & \nod    & \nod & 13,18     \\
060502b & 0.09    & $(4.0\pm 0.5)\times 10^{-8}$ & \ra{18}{35}{45.53} & \dec{+52}{37}{52.9} & 3.7 & N & \nod   & $25.83\pm 0.05$ & \nod & \nod    & \nod & 13,19--21 \\
060801  & 0.50    & $(8.1\pm 1.0)\times 10^{-8}$ & \ra{14}{12}{01.35} & \dec{+16}{58}{53.7} & 2.4 & N & 1.1304 & $22.97\pm 0.11$ & 0.6  & 6.1     & \nod & 13,22--24 \\
061006  & 0.42    & $(1.4\pm 0.1)\times 10^{-6}$ & \ra{07}{24}{07.66} & \dec{-79}{11}{55.1} & 0.5 & Y & 0.4377 & $22.65\pm 0.09$ & 0.1  & 0.2     & 8.6  & 13,25--27 \\
061201  & 0.8     & $(3.3\pm 0.3)\times 10^{-7}$ & \ra{22}{08}{32.09} & \dec{-74}{34}{47.1} & 0.2 & Y & \nod$^e$ & \nod          & \nod & \nod    & \nod & 28        \\
061210  & 0.19    & $(1.1\pm 0.2)\times 10^{-6}$ & \ra{09}{38}{05.27} & \dec{+15}{37}{17.3} & 1.8 & N & 0.4095 & $21.00\pm 0.02$ & 0.9  & 1.2     & 8.8  & 13,29     \\ 
061217  & 0.21    & $(4.6\pm 0.8)\times 10^{-8}$ & \ra{10}{41}{39.32} & \dec{-21}{07}{22.1} & 3.8 & N & 0.8270 & $23.33\pm 0.07$ & 0.4  & 2.5     & \nod & 13,30     \\ 
070429b & 0.5     & $(6.3\pm 1.0)\times 10^{-8}$ & \ra{21}{52}{03.84} & \dec{-38}{49}{42.4} & 6.2 & N & 0.9023 & $23.22\pm 0.10$ & 0.6  & 1.1     & \nod & 31        \\ 
070707  & 1.1     & $(1.4\pm 0.2)\times 10^{-6}$ & \ra{17}{50}{58.55} & \dec{-68}{55}{27.2} & 0.5 & Y & \nod   & $>24.4$         & \nod & \nod    & \nod & 32--33    \\ 
070714a & 2.0     & $(1.5\pm 0.2)\times 10^{-7}$ & \ra{02}{51}{43.10} & \dec{+30}{14}{34.2} & 3.2 & N & \nod   & \nod            & \nod & \nod    & \nod & 34        \\ 
070714b & 3$^f$   & $(7.2\pm 0.9)\times 10^{-7}$ & \ra{03}{51}{22.30} & \dec{+28}{17}{50.8} & 0.4 & Y & 0.9230 & $24.92\pm 0.23$ & 0.1  & 0.4     & \nod & 31,35--36 \\
070724  & 0.4     & $(3.0\pm 0.7)\times 10^{-8}$ & \ra{01}{51}{13.96} & \dec{-18}{35}{40.1} & 2.2 & N & 0.4571 & $20.53\pm 0.03$ & 1.4  & 2.5     & 8.9  & 37--39    \\
070729  & 0.9     & $(1.0\pm 0.2)\times 10^{-7}$ & \ra{03}{45}{16.04} & \dec{-39}{19}{19.9} & 2.5 & N & \nod   & $23.32\pm 0.23$ & \nod & \nod    & \nod & 40--41    \\
070809  & 1.3     & $(1.0\pm 0.1)\times 10^{-7}$ & \ra{13}{35}{04.41} & \dec{-22}{08}{28.9} & 4.8 & N & \nod   & $24.86\pm 0.27$ & \nod & \nod    & \nod & 42--43    \\
071227  & 1.8     & $(2.2\pm 0.3)\times 10^{-7}$ & \ra{03}{52}{31.26} & \dec{-55}{59}{03.5} & 0.3 & Y & 0.3940 & $20.54\pm 0.03$ & 1.2  & \nod    & \nod & 44--46    \\
080123  & 0.4     & $(5.7\pm 1.7)\times 10^{-7}$ & \ra{22}{35}{46.10} & \dec{-64}{54}{03.2} & 2.1 & N & \nod   & \nod            & \nod & \nod    & \nod & 47       
\enddata
\tablecomments{Properties of the short GRBs and their host galaxies
discussed in this paper, including (i) GRB name, (ii) duration, (iii)
fluence, (iv-vi) position of the X-ray or optical afterglow including
uncertainty, (vii) whether an optical afterglow was detected, (viii)
spectroscopic redshift, (ix) host galaxy $R$-band magnitude corrected
for Galactic extinction \citep{sfd98}, (x) rest-frame $B$-band
luminosity, (xi) star formation rate, (xii) metallicity, and (xiii)
relevant references.\\
$^a$ All XRT positions are from the catalogs of \citet{but07} and
\citet{gtb+07}.\\
$^b$ Corrected for Galactic extinction \citep{sfd98}.\\
$^c$ The light curve is dominated by a 0.25 s hard spectrum spike,
with a BATSE duration of $T_{90}=1.3$ s.\\
$^d$ The XRT error circle contains a single galaxy with red optical
and near-IR colors indicative of an early-type galaxy at a photometric
redshift of $\sim 1.8$ \citep{ber06}.  However, no spectroscopic
confirmation is available.\\ 
$^e$ The nature of the host galaxy remains unclear, given the location
of an Abell cluster at $z=0.0865$ about $8'$ away \citep{gcn5995}, a
galaxy at $z=0.111$ about $17''$ away \citep{gcn5952,sdp+07}, and a
faint galaxy of unknown redshift $\lesssim 1''$ away.  All three
associations have a statistical significant of about $10\%$.  We note
that the galaxy at $z=0.111$ has a specific star formation rate of 1.2
M$_\odot$ yr$^{-1}$ $L_*^{-1}$ \citep{sdp+07}, in good agreement with
the host sample presented in this paper.  In addition, we measure for
this galaxy a luminosity of $M_B\approx -19.1$ mag and a metallicity
of $12+{\rm log(O/H)}=8.8\pm 0.2$, in excellent agreement with the
luminosity-metallicity relation.\\
$^f$ The light curve is dominated by a short and hard initial spike,
with a small spectral lag \citep{gcnr70}.\\
References: [1] \citet{gso+05}; [2] \citet{bpp+06}; [3]
\citet{vlr+05}; [4] \citet{ffp+05}; [5] \citet{hwf+05}; [6]
\citet{bcb+05}; [7] \citet{bpc+05}; [8] \citet{gbp+06}; [9]
\citet{fsk+07}; [10] \citet{ber06}; [11] \citet{pbc+06}; [12]
\citet{lmf+06}; [13] \citet{bfp+07}; [14] \citet{bgc+06}; [15]
\citet{sbk+06}; [16] \citet{gcn4550}; [17] \citet{gcn4565}; [18]
\citet{rvp+06}; [19] \citet{gcn5064}; [20] \citet{gcn5093}; [21]
\citet{bpc+07}; [22] \citet{gcn5381}; [23] \citet{gcn5381}; [24]
\citet{gcn5389}; [25] \citet{gcn5704}; [26] \citet{gcn5710}; [27]
\citet{gcn5723}; [28] \citet{sdp+07}; [29] \citet{gcnr20}; [30]
\citet{gcnr21}; [31] \citet{cbn+08}; [32] \citet{gcn6607}; [33]
\citet{gcn6613}; [34] \citet{gcn6622}; [35] \citet{gcn6623}; [36]
\citet{gfl+08}; [37] \citet{gcn6656}; [38] \citet{gcn6661}; [39]
\citet{gcn6665}; [40] \citet{gcn6678}; [41] \citet{gcn6680}; [42]
\citet{gcnr80}; [43] \citet{gcn6774}; [44] \citet{gcn7148}; [45]
\citet{gcn7152}; [46] \citet{gcn7154}; [47] \citet{gcnr111}.}
\end{deluxetable}

\clearpage
\begin{deluxetable}{llllllcccl}
\tabletypesize{\footnotesize}
\tablecolumns{10}
\tabcolsep0.05in\footnotesize
\tablewidth{0pc}
\tablecaption{Emission Line Fluxes and Metallicity Indicators
\label{tab:lines}}
\tablehead {
\colhead {GRB}                 &
\colhead {$F_{\rm [OII]}$}     &
\colhead {$F_{\rm [OIII]}$}    &
\colhead {$F_{\rm H\beta}$}    &
\colhead {$F_{\rm H\alpha}$}   &
\colhead {$F_{\rm [NII]}$}     &
\colhead {${\rm log}(R_{23})$} &
\colhead {${\rm log}(O_{32})$} &
\colhead {$12+{\rm log}({\rm O/H})\,^a$} &
\colhead {Refs.}               \\\cline{2-6} 
\colhead {}                    &
\multicolumn{5}{c}{($10^{-17}$ erg cm$^{-2}$ s$^{-1}$)} &
\colhead {}                    &                   
\colhead {}                    &                   
\colhead {}                    &                   
\colhead {}                    
}
\startdata
061006  & 4.1  & 1.5  & 0.9 & \nod & \nod & 0.79 & $-0.44$ & 8.63 & This paper \\
061210  & 22   & 10.5 & 7.6 & 11   & 2.4  & 0.63 & $-0.32$ & 8.82 & This paper \\
070724  & 37   & 17   & 15  & \nod & \nod & 0.56 & $-0.34$ & 8.88 & This paper \\\hline
050709  & \nod & 26   & 6.6 & 26   & 1.8  & \nod & \nod    & 8.50 & 1 \\
051221a & 10.3 & 5.9  & 3.9 & \nod & \nod & 0.62 & $-0.24$ & 8.84 & 2
\enddata
\tablecomments{Emission line fluxes and metallicity indicators.  The
quantities $R_{23}$ and $O_{32}$ are defined in \S\ref{sec:metal}.\\
$^a$ The typical uncertainty on the metallicity is about $0.1$ dex, 
dominated by the systematics in the $R_{23}$ and $O_{32}$ calibrations 
\citep{kk04}.\\
References: [1] \citet{ffp+05}; [2] \citet{sbk+06}.}
\end{deluxetable}

\clearpage

\clearpage

\clearpage

\clearpage

\end{document}